\documentclass[letter,preprint,preprintnumbers,nofootinbib]{revtex4-1}
\usepackage{amsfonts,amssymb,amsmath,color,graphicx}
\usepackage{caption}
\usepackage{subcaption}
\begin{document}

\title{Inflection Points and the Power Spectrum}

\author{Sean Downes and Bhaskar Dutta}

\affiliation{Department of Physics and Astronomy, Texas A\&M University, College Station, TX 77843-4242, USA}

\begin{abstract}
Inflection point inflation generically includes a deviation from slow-roll when the inflaton approaches the inflection point. Such deviations are shown to be generated by transitions between singular trajectories. The effects on the power spectrum are studied within the context of universality classes for small-field models. These effects are shown to scale with universality parameters, and can explain the anomalously low power at large scales observed in the CMB. The reduction of power is related to the inflection point's basin of attraction. Implications for the likelihood of inflation are discussed.

\end{abstract}
\noindent MIFPA-12-39 \\ November 2012

\maketitle

\tableofcontents
\newpage
\section{Introduction}

Inflation is a leading theoretical explanation for the origin of our universe\cite{Guth:1980zm,Linde:1981mu,Albrecht:1982wi}. In addition to resolving many longstanding problems with big bang cosmology, it also motivates the observed, nearly scale invariant spectrum of density perturbations. Such primordial perturbations are widely believed to have seeded the formation of structure, which is still being uncovered at the largest scales. As a possible explanation of these perturbations, inflation also offers a quantitative description of the observed temperature anisotropies in the cosmic microwave background (CMB). 

Inflation utilizes a scalar field to drive exponential expansion of the spacetime metric. Quantum perturbations of an accelerating background are violently stretched to cosmic scales, where they exit the particle horizon, become nondynamical and wait for the horizon to catch up. The once-quantum fluctuations reenter the horizon as a random gravitation potential for the constituents of the cosmic fireball --- the aftermath of the big bang. The density perturbations can therefore be seen today as thermal fluctuations in the CMB. It is the detailed study of these ancient photons that has made the case for inflation particularly compelling\cite{Komatsu:2010fb}.

Despite being a successful theoretical paradigm, inflation has resisted a concrete embedding into known physical theories. At present, it seems that neither string theory nor particle physics have any distinguished role or \textit{a priori} need for inflation. Yet there are dozens of incarnations of inflation, each manifested in hundreds of models. Inflation, it would seem, stands embarrassed by its riches. Its agent --- the hypothetical scalar field dubbed the ``inflaton'' --- insists on remaining anonymous.

In the past decade there have been attempts to uncover its identity. One major approach involves search for nongaussianities in the CMB \cite{Maldacena:2002vr,Babich:2004gb}, formalized through the so-called effective field theory of inflation\cite{Cheung:2007st}. The idea being that nonlinear effects in the CMB can be tied to nonlinear terms in the scalar potential, which may shed light on the inflaton's identity. 

A complimentary  approach was recently proposed in \cite{Downes:2011gi}. Here, model independent details of how inflation embeds into models derived from string theory and particle physics were codified in a set of universality classes. For the inflection point inflation scenario, this universal behavior was shown to relate phenomena as disparate as density perturbations and supersymmetry breaking. The theoretical technology afforded by V.~I.~Arnold's Singularity Theory \cite{arnold} gave rise to scaling relations between physical observables.

In this work we investigate the effect of such scaling relations on the linear perturbations. We shall see that universality works in concert with attractor dynamics \cite{Downes:2012xb} to leave an observable footprint on the angular power spectrum of the CMB. In particular, we shall see how the inflaton temporarily leaves slow roll, giving rise to an observable reduction in power.

Put a different way, nontrivial dynamics of the inflaton can reduce the power at scales where such modes left the horizon. This leads to definite observational consequences. Indeed, at the largest scales accessible there is a sharp, anomalous decrease in power in the CMB. Possible origins for these effects have been studied previously \cite{Contaldi:2003zv,Schwarz:2009sj,Cline:2003ve,Ramirez:2012gt,Dudas:2012vv}, but generally in the context of the minimal ``just enough'' inflation. This work substantially generalizes this approach.

Taken at face value, these considerations are speculative. The azimuthal modes of the power spectrum are taken to be independent measurements. Therefore, large scale harmonic modes means low statistics, and therefore a theoretical limit on the precision of the angular power spectrum. This uncertainty, known as \textit{cosmic variance}, can accommodate the anomalously low power at large angles. Nevertheless, one may hope that the same physics which yields a thermal power spectrum with low power at large scales may also render a better global fit to the cosmological data. 

The remainder of this work is organized as follows: in Sec.~\ref{sec:two} we review a few relevant features from slow roll inflation. In Sec.~\ref{sec:three} we describe the systematic analysis of the power spectrum and present the results. In Sec.~\ref{sec:four} we conclude.

\section{Selections from Slow-Roll Inflation}\label{sec:two}
We begin our discussion by developing a few relevant details from slow-roll inflation. First, we develop the attractor dynamics using the formalism in \cite{Downes:2012xb}. Next, we discuss the conditions for the inflaton trajectory to temporarily fall out of slow-roll. In particular, we emphasize the transition from chaotic to inflection point inflation. Finally, we review the ``ensemble of universes'' given by the set of all possible trajectories of the gravity-scalar system; it is the framework for performing our systematic analysis of the power spectrum in Section~\ref{sec:three}. This is the basin of attraction in phase space; we demonstrate how it varies with the catastrophe parameters of \cite{Downes:2011gi}.

\subsection{Slow-roll and other singular trajectories}\label{sec:slowroll}
We work in natural units, with $c=\hbar=M_P=1/\sqrt{8\pi G}=1$. The field equations which govern the Fridemann-Lema\^itre-Robertson-Walker-scalar system are given by,
\begin{eqnarray}\label{oldeqna}\ddot{\phi}+3H\dot{\phi}+V_{\phi}&=&0,\\-3H^2 +\frac{1}{2}\dot{\phi}^2 + V &=& 0,\\ \dot{H}+\frac{1}{2}\dot{\phi}^2 &=& 0.\end{eqnarray}
A subscript $\phi$ denotes a partial derivative with respect to $\phi$. Dots corresponds to time derivatives.
To separate these coupled equations and explicitly manifest the attractor dynamics, we parametrize time with the (suitably normalized) scalar factor,
$$t\rightarrow N = \log a/a_0,$$
which leads to relations like
$$\dot{\phi} = H\phi^{\prime},$$
where primes denote derivatives with respect to $N$. The normalization $a_0$ is the value of the scale factor today.

With this parametrization, the field equations can be rewritten,
\begin{eqnarray}\label{neweqna}\phi^{\prime\prime}&=&-3\left(1-\frac{1}{6}\phi^{\prime 2}\right)\left(\phi^{\prime}+\frac{V_{\phi}}{V}\right),\\ \label{neweqnb}3H^2 &=& \frac{V}{1-\frac{1}{6}\phi^{\prime 2}},\\ \label{neweqnc}H^{\prime} &=& -\frac{1}{2}\phi^{\prime 2} H.\end{eqnarray}

The field equation \eqref{neweqna} admits three ``singular'' trajectories were the right hand side vanishes. The two solutions, $\phi^{\prime} = \pm \sqrt{6}$, corresponds to kinetic domination of the energy density. These are ``fast-roll'' solutions and are typically dynamical repulsors. The other, ``slow-roll'' trajectory --- call it $\phi_{\star}$ --- solves,
\begin{equation}\label{slowroll}\phi^{\prime} = -\frac{V_{\phi}}{V}.\end{equation}
This is only an approximate solution, since
\begin{equation}\label{srapprox}\phi^{\prime\prime}_{\star} = -\phi^{\prime}_{\star}\left[\frac{V_{\phi\phi}}{V} - \left(\frac{V_{\phi}}{V}\right)^2\right]_{\phi\,=\,\phi_{\star}},\end{equation}
which does not vanish in general. However, the condition that $\phi_{\star}$ approximate a solution to \eqref{neweqna}  --- that $\phi^{\prime\prime}_{\star}$ nearly vanishes --- agrees with the familiar slow roll conditions,

\begin{equation}\label{oldslowroll}\frac{1}{2}\left(\frac{V_\phi}{V}\right)^2\ll1,\quad \frac{V_{\phi\phi}}{V}\ll 1.\end{equation}

Indeed, the vanishing of \eqref{srapprox} is the more precise statement of \eqref{oldslowroll}.

Despite the failure of $\phi_{\star}$ to solve the field equation, it is still a dynamical attractor, at least so long as $\phi_{\star}$ is less than $\sqrt{6}$. The details of the attractor/repulsor trajectories are understood \cite{Salopek:1990jq,1994PhRvD..50.7222L}, and were reviewed in this framework in \cite{Downes:2012xb}. 

The small deviations from a full solution to \eqref{slowroll} has direct consequences for those $\phi$ that do solve \eqref{neweqna}. The resulting deviation from $\phi_{\star}$ leads to consequences in the power spectrum, and is the principal focus of this work.

\subsection{Deviations from slow-roll}
A convenient parametrization of the deviation from the slow-roll trajectory $\phi_{\star}$ is given by the variable $\Xi$,
$$\Xi = \frac{\phi^{\prime\prime}}{\phi^{\prime}}.$$
Simple algebra reveals that,
\begin{equation}\label{xi}\Xi = \frac{1}{2}(\phi^{\prime}+\sqrt{6})(\phi^{\prime}-\sqrt{6})(1-\frac{V_{\phi}}{\phi^{\,\prime}\,V}).\end{equation}
So $\Xi$ vanishes for all ``singular'' solutions of \eqref{neweqna}, both slow-roll and fast-roll. This is important to bear in mind, as $\Xi$ also enters directly into the mode equations for the linear perturbations, 
\begin{equation}\label{pertsa}u_{k}^{\prime\prime} + \left(1-\frac{1}{2}\phi^{\prime 2}\right)u_k^{\prime} + \left[\left(\frac{k}{aH}\right)^2 + (1+\Xi)\left(\Xi + \frac{1}{2}\phi^{\prime 2}\right) - \Xi^{\prime}\right]u_k=0,\end{equation}
and can therefore impact the power spectrum when it is of order unity. Quantifying this impact with respect to the scaling phenomena observed in \cite{Downes:2011gi} is the central focus of this paper. We will discuss this and \eqref{pertsa} in more detail in the next section.

As $\phi$ transitions between different singular trajectories, $\Xi$ can become large. To build an intuition for such effects, we consider three examples of large $\Xi$: when the $\phi$ starts from rest, when $\phi$ starts from a fast-roll trajectory, and when it transitions from chaotic to inflection point inflation. All three can may lead to observable effects on the power spectrum. This has been studied, for instance, in \cite{Contaldi:2003zv,Schwarz:2009sj,Cline:2003ve}.

\subsubsection*{Field velocity and $\Xi$}

\begin{figure}[h]
    \includegraphics[width=0.6\textwidth]{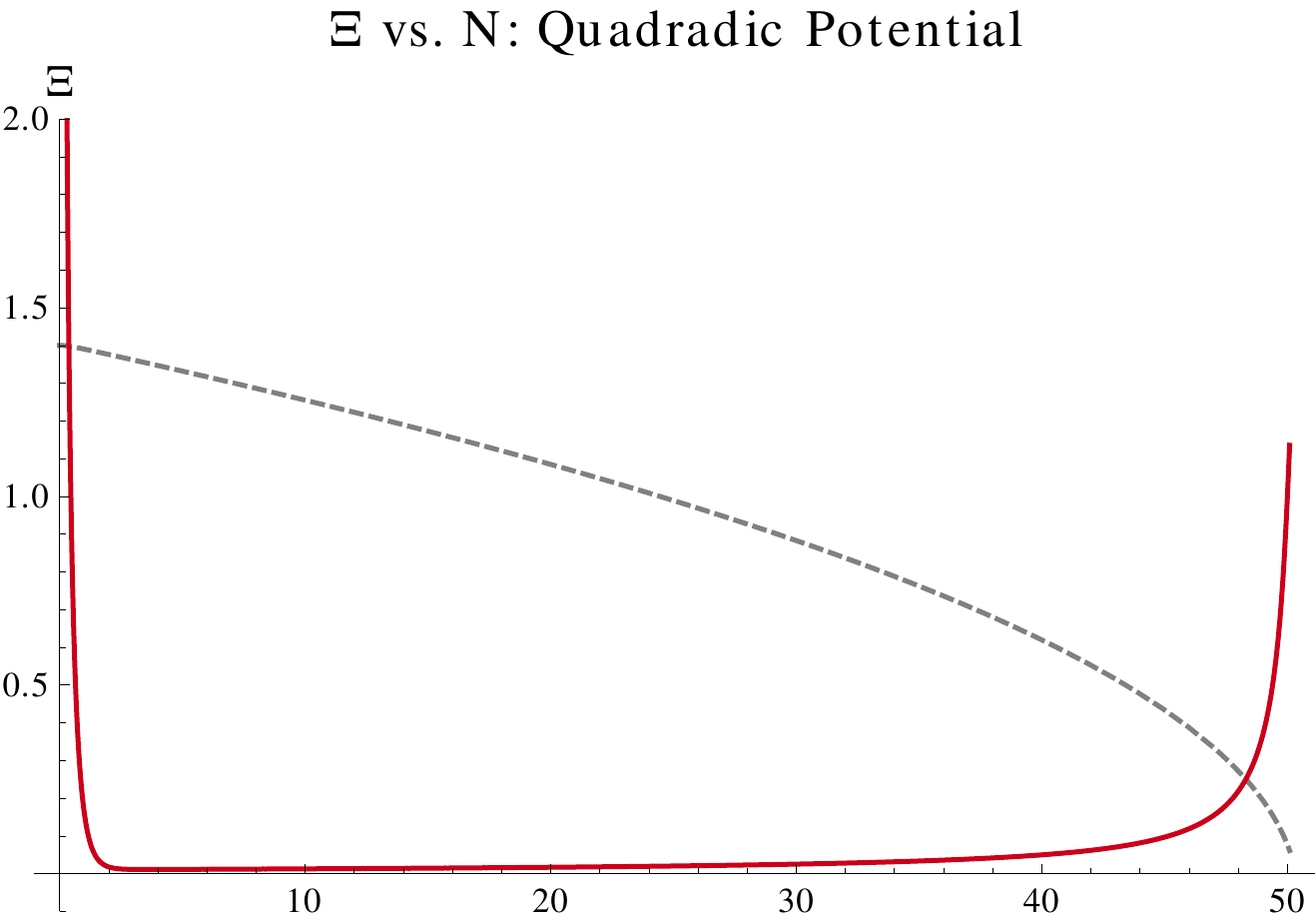}
  \caption{$\Xi$ as a function of $N$ for chaotic inflation on a quadratic potential (solid red line). $\phi$ starts from rest and is shown (dashed, with arbitrary units) for qualitative comparison. Near $N=50$, slow-roll ends and $\phi$ accelerates towards the minimum.}
\label{fig:xichaos}
\end{figure}

We begin with the chaotic inflation scenario. Consider a field slowly rolling in a quadratic potential, $V\propto \phi^2$. If $\phi$ starts at rest, it must first accelerate towards the slow-roll trajectory, \eqref{slowroll}. Thus, a nearly vanishing field velocity rarely qualifies as slow-roll. Indeed, if the field starts from rest, $\Xi$ can be quite large as the field rapidly accelerates toward the slow-roll trajectory. When $\phi^{\prime}$ vanishes, $\Xi$ diverges. This is illustrated numerically in Fig.~\ref{fig:xichaos}.

\begin{figure}[h]
    \includegraphics[width=0.6\textwidth]{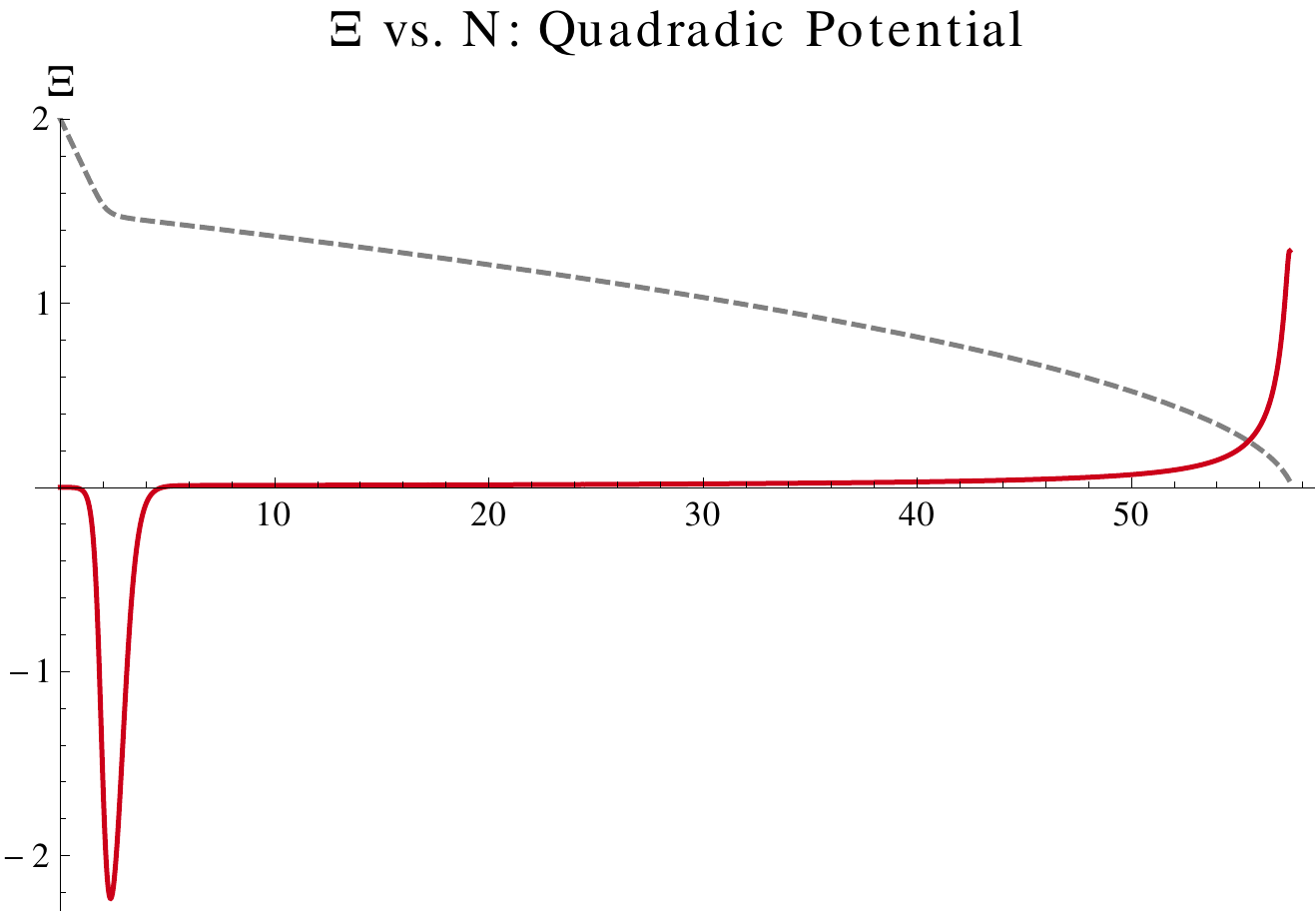}
  \caption{$\Xi$ as a function of $N$ for chaotic inflation on a quadratic potential (solid red line). $\phi$ starts from fast roll, and is shown (dashed, with arbitrary units) for qualitative comparison. $\Xi$ spikes as the field transitions from fast-roll to slow-roll. $\Xi$ spikes again as $\phi$ accelerates towards the minimum of the potential.}
\label{fig:xifast}
\end{figure}

Alternatively, the inflaton may start near the fast-roll condition, $\phi^{\prime}\approx -\sqrt{6}$. In this case, $\Xi$ starts off nearly vanishing, but spikes when the field transitions to slow-roll, as shown in Fig.~\ref{fig:xifast}.

These two examples involved chaotic inflation. There are other cases of interest, like a transition between chaotic and inflection point inflation.

\subsubsection*{Transition from chaotic to inflection point inflation}

Chaotic inflation solves the slow roll conditions \eqref{oldslowroll} with large $V$, which necessitates large field excursions. Inflection point inflation satisfies \eqref{oldslowroll} by a vanishingly small $V_{\phi}$ and $V_{\phi\phi}$. In this sense, inflection point inflation is a misnomer. $V$ must possess a degenerate critical point; its first derivative must also vanish at an inflection point. Inflation then occurs over a small field excursion in the vicinity of the degenerate critical point.

Since the first two derivatives vanish, the Taylor expansion near the ``inflection point'' has the form,
\begin{equation}\label{prototype}V \approx V_0 + V_3\phi^3 + \mathcal{O}(\phi^4).\end{equation}

\begin{figure}[h]
    \includegraphics[width=0.6\textwidth]{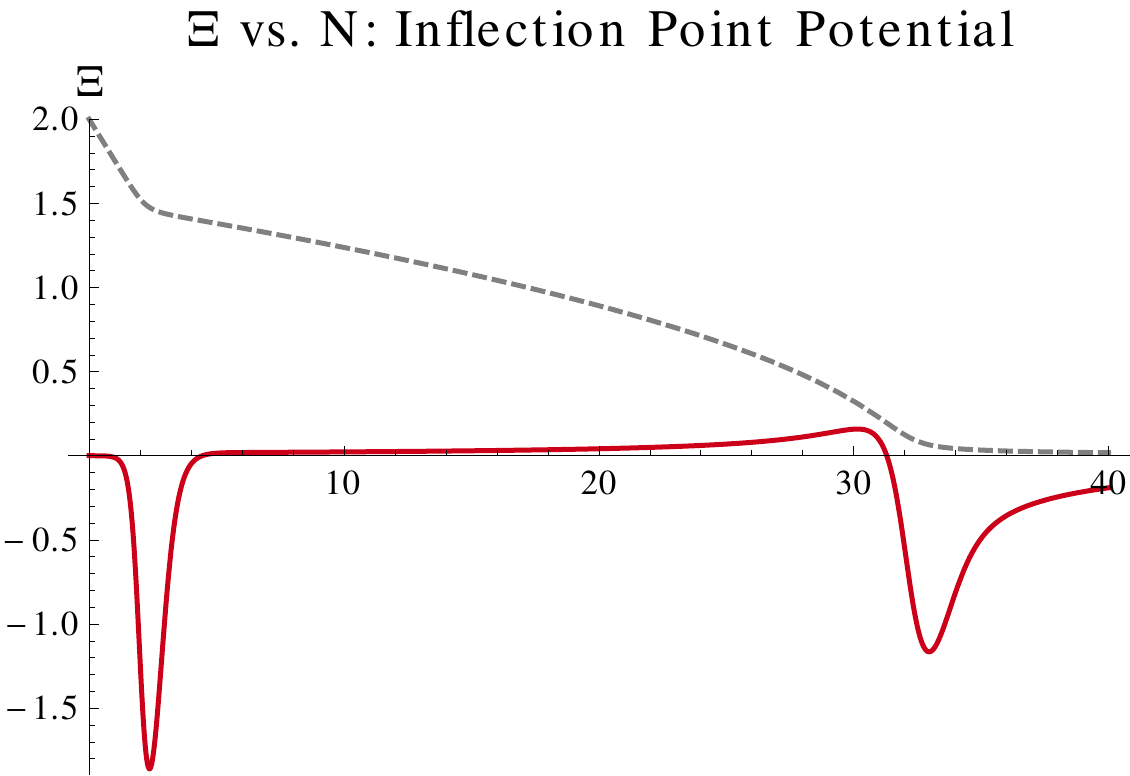}
  \caption{$\Xi$ as a function of $N$ for an inflection point potential (solid red line). $\phi$ starts from fast roll, and is shown (dashed, with arbitrary units) for qualitative comparison.}
\label{fig:xiinflect}
\end{figure}

These two inflationary scenarios are not mutually exclusive. As a result of attractor dynamics, reviewed in the next subsection, the inflaton can transition from a period of chaotic to inflection point inflation. As illustrated in Fig.~\ref{fig:xiinflect}, $\Xi$ spikes as $\phi$ approaches the inflection point. This transition from chaotic to inflection point potentials was first studied in \cite{Itzhaki:2008hs} and later in \cite{Downes:2011gi}. Physically, trajectories close to slow-roll in the chaotic regime must brake abruptly at the inflection point. The ``braking'' in $\phi_{\star}$ is too strong to be physical, leading instead to a spike in $\Xi$. It is an example of the failure of an exact solution $\phi$ to match $\phi_{\star}$ mentioned in Sec.~\ref{sec:slowroll}.  

\begin{figure}[h]
    \includegraphics[width=0.6\textwidth]{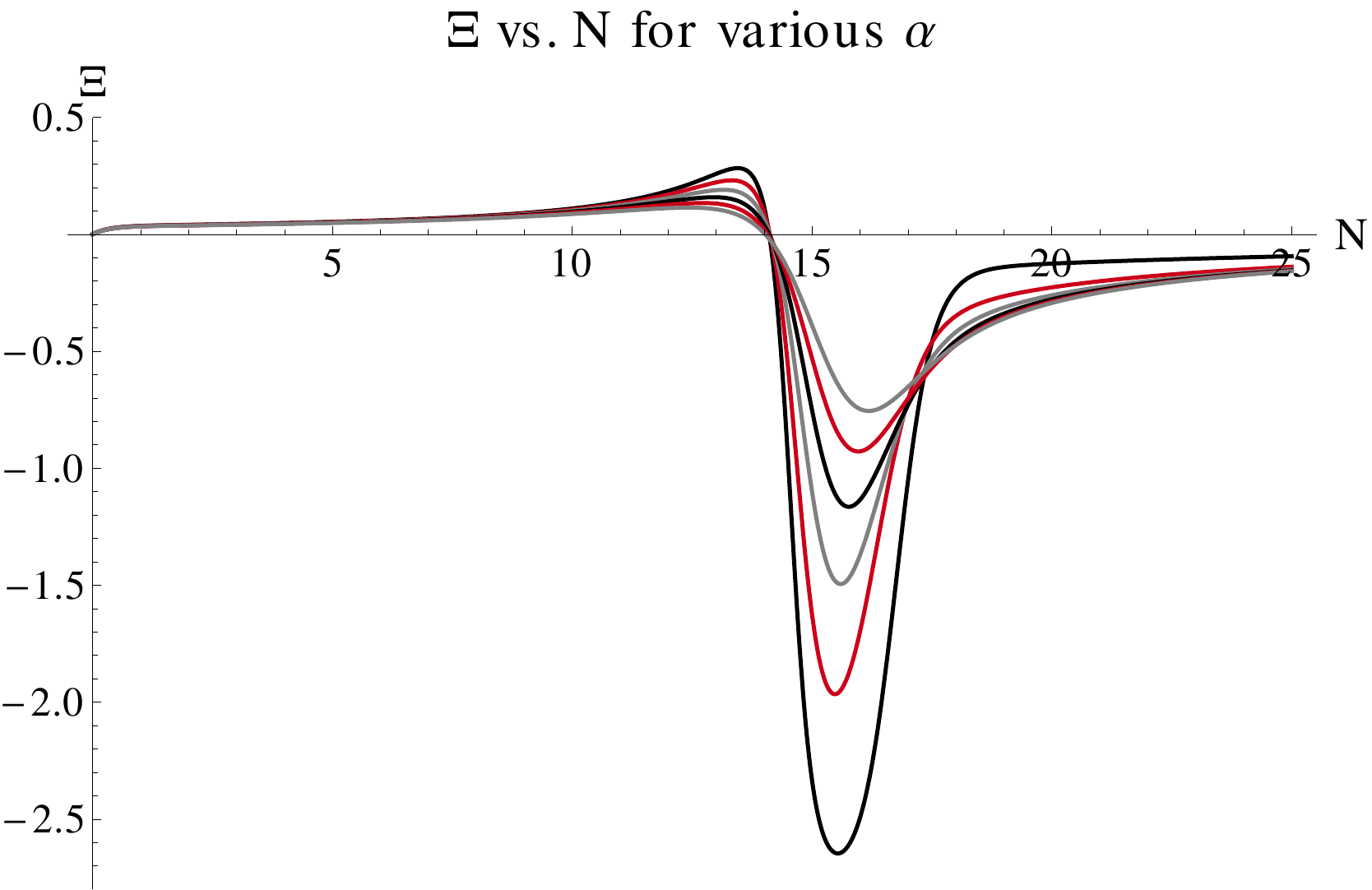}
  \caption{$\Xi$ plotted for trajectories with various values of the cubic coupling $\alpha$. Each was started with the same, slow-roll boundary conditions. As $\alpha$ increases, the amplitude of $\Xi$ rapidly decreases. Here $\alpha$ ranges from 0.7 to 1.2 in increments of 0.1.}
\label{fig:inflection}
\end{figure}

The cubic coupling at the inflection point is the leading term near the critical point, as seen in \eqref{prototype}. This coupling strongly influences $\Xi$. As seen in Fig.~\ref{fig:inflection}, the amplitude of $\Xi$ decreases inversely to this coupling. In inflection point models, this is the most important contribution to $\Xi$.

Since $\Xi$ can be quite large as the field approaches the inflection point, there is the possibility for an observable effect in the power spectrum. We investigate this in the next section. Before that, we lay the groundwork for a systematic analysis by discussing the inflection point's basin of attraction.

\subsection{The basin of attraction}\label{sec:basin2}

The attractor dynamics of slow-roll inflation are well known, particularly in the chaotic inflation scenario \cite{Linde:2007fr,kofman}. What is less well known is that inflection points are very efficient attractors \cite{Allahverdi:2008bt,Downes:2012xb}. We review the basin of attraction for the simplest class of inflection point models; the generalization is straight forward.

Models of inflection point inflation fall into universality classes \cite{Downes:2011gi}, depending on the number of parameters in the potential. The canonical representative of the class of two-parameter models $(A_3)$ is,
$$V = \frac{1}{4}\phi^{4}+\frac{1}{2}a\phi^{2}+b\phi + {\rm constant}.$$
The condition on the parameters for a degenerate critical point (``inflection point'') is
\begin{equation}\label{cusp}\left(\frac{a}{3}\right)^3 + \left(\frac{b}{2}\right)^2 = 0.\end{equation}
This gives an inflection point at $\phi=\alpha$, and a minimum --- a vacuum suitable for reheating --- at $\phi=-3\alpha$.  After shifting the origin of field space to coincide with the inflection point, what remains is a family of potentials, parametrized by $\alpha$.
\begin{equation}\label{A3pot}V = \frac{1}{4}\phi^4 + \alpha \phi^3 + \frac{27}{4}\alpha^4.\end{equation}
Thus, for two-parameter models of inflection point inflation, a solution is specified by a choice of $\alpha$ and an initial point in phase space.

\begin{figure}
	\centering
         \includegraphics[width=0.55\textwidth]{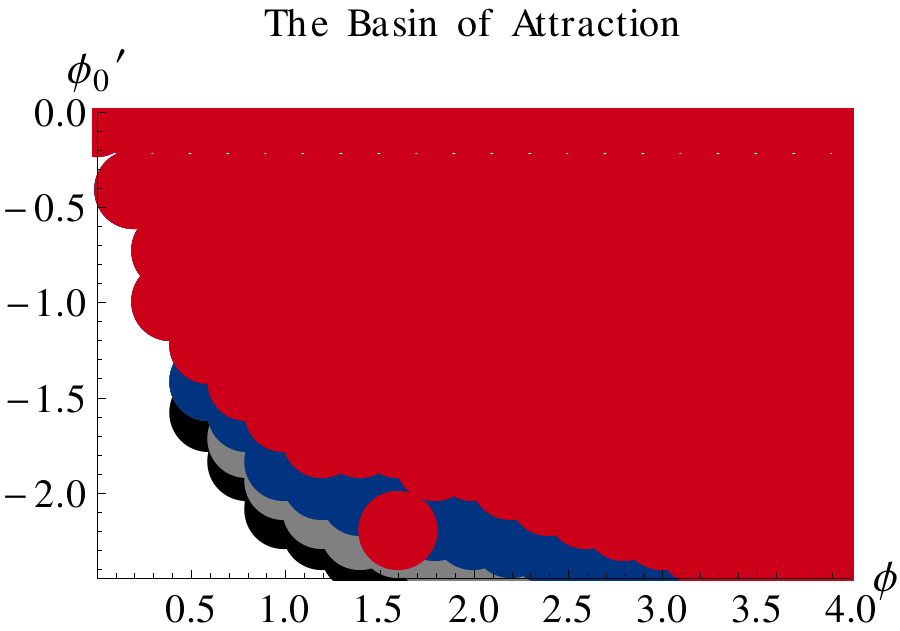}	

        \caption{Initial conditions for the scalar field subject to the potential \eqref{A3pot}. Points in the shaded regions come to rest at the inflection point, others overshoot. The basin of attraction shrinks with smaller $\alpha$. The colored bands correspond to different values of $\alpha$: 2,1,0.8 and 0.7.}\label{fig:basin}
\end{figure}

For any choice of three such numbers, some solutions will asymptotically approach the inflection point and come to rest. Others will overshoot. Owing to the solitonic nature of these trajectories, the boundary of the basin of attraction in this space is a transcendental function. Despite this, such an analysis can be carried out numerically \cite{Downes:2012xb}. For fixed $\alpha$, the basin of attraction can be seen in Fig~\ref{fig:basin}.

\subsubsection*{Finite e-foldigs}
So far we have studied idealized potentials with exactly degenerate critical points. The attractor behavior gives rise to infinitely many e-foldings within the basin of attraction described. Physically, inflation must end, so the critical points must not be precisely degenerate. As discussed is \cite{Downes:2012xb}, this is good, otherwise the subset of viable couplings would be of measure zero.

Inflation ends due to the presence of an additional, small term in the potential. In \eqref{cusp}, this means a slight shift in $b$. This can be modelled by a slightly deformed potential,
\begin{equation}\label{A3potdefo}V = \frac{1}{4}\phi^4 + \alpha\phi^3 + \lambda_1\phi + \frac{27}{4}\alpha^4+ \mathcal{O}(\lambda_1).\end{equation}
The constant term has been deformed slightly to keep the minima approximately zero. The number of e-foldings associated to the inflection point is approximately \cite{Linde:2007jn},
\begin{equation}\label{efolds}N_e \approx \frac{\pi}{2\sqrt{3\alpha \lambda_1}}.\end{equation}
Note that arbitrarily many e-foldings of slow-roll inflation may exist prior to the contribution of inflection point. Physically, of course, we are interested in the the point where the observed perturbations in the CMB entered the horizon - last sixty or so e-foldings.

So long as $\lambda_1 \ll \alpha^3$, as is typically required to achieve sufficiently many e-foldings to match observations, the basins of attraction discussed above reasonably models the physically viable solutions. In the following section, we shall focus on solutions within this basin.

\subsubsection*{Critical Couplings}\label{sec:estadocritical}
We close this section with a technical aside. Once again, we consider the case where $\lambda$ vanishes. 

Not all values of $\alpha$ in \eqref{A3pot} give rise to a finite basin of attraction. Below a critical value, $\alpha_C\sim 0.66$, the basin is simple a point; overshooting the inflection point is inevitable. This was first observed in the one-parameter ($A_2$) case \cite{Itzhaki:2008hs}, and was generalized in \cite{Downes:2012xb} which also mapped out the basin of attraction. In the next section we shall exclusive focus on couplings well inside the basin, so that we may sensibly talk about sufficient inflation. Despite that, it is worth pausing to mention the interesting behavior near these critical couplings.

\begin{figure}[h]
  \centering
    \includegraphics[width=0.6\textwidth]{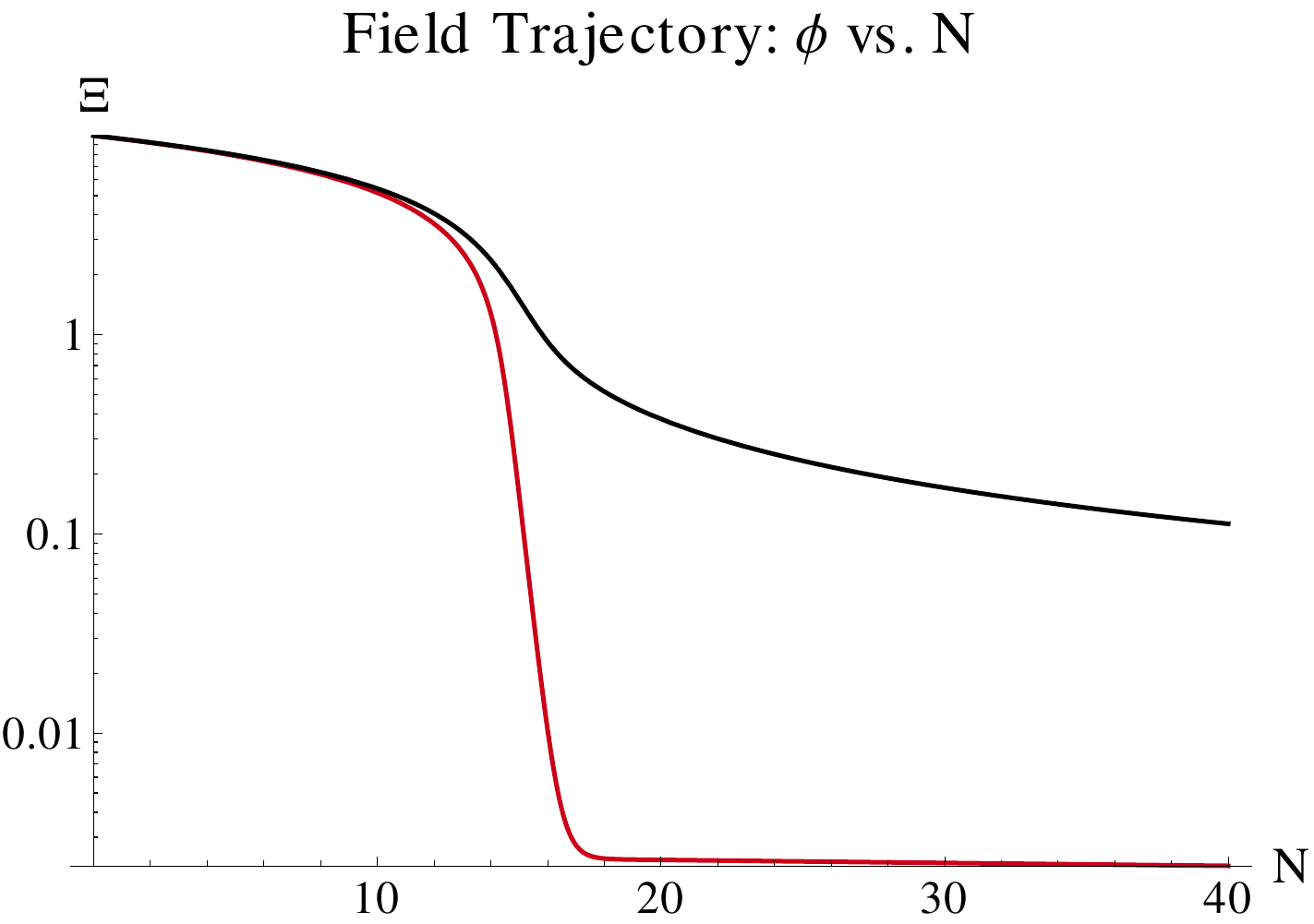}
  \caption{Trajectories of $\phi$ started on slow-roll for the potential \eqref{A3pot}. The blue curve uses $\alpha=0.6599$ and the purple uses $\alpha=1.1599$. Note how the blue curve abruptly approaches the inflection point, whereas the purple curve is slowly asymptoting towards it.}
\label{fig:trajectory}
\end{figure}

We saw in Fig.~\ref{fig:inflection} that $\alpha$ decreases, the amplitude of deviation $\Xi$ grows. This deviation occurs as the field brakes while approaching the inflection point. As $\alpha$ approaches $\alpha_C$ from above, the required braking becomes substantial. This is depicted in Fig.~\ref{fig:trajectory}.

\begin{figure}[h]
  \centering
    \includegraphics[width=0.6\textwidth]{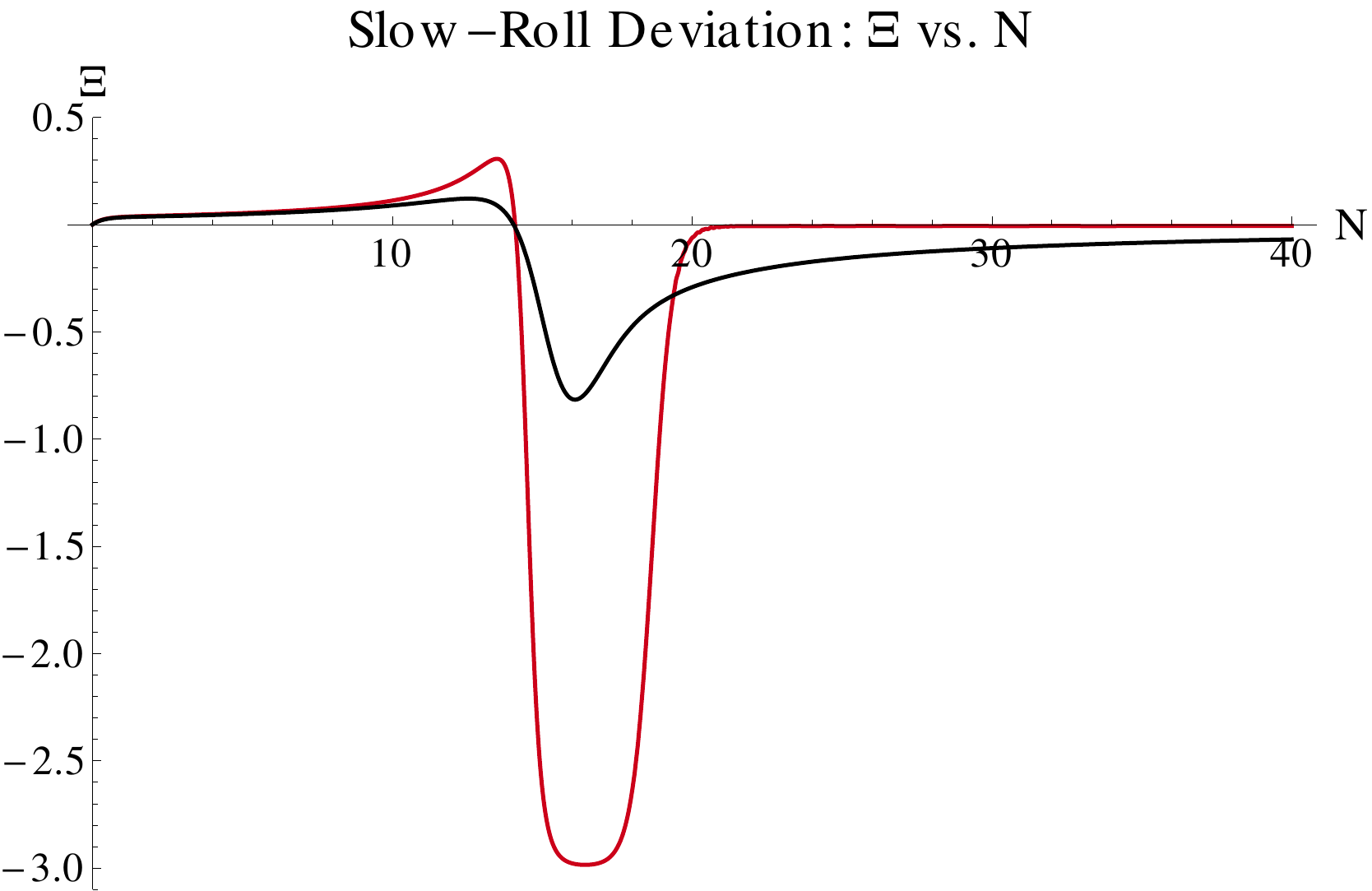}
  \caption{The deviation from slow-roll, $\Xi$, is plotted versus $N$. The blue curve uses $\alpha=0.6599$ and the purple, representative of those in Fig.~\ref{fig:inflection}, uses $\alpha=1.1599$. Note how the blue curve reaches a lower bound.}
\label{fig:deviation}
\end{figure}

With significant breaking comes significant deviation from slow-roll. The deviation finds a lower bound\footnote{The boundedness in this context relies on the monotonic approach of the field to the inflection point. Curiously, it can be shown that the $-3$ is precisely the coefficient of the right-hand side of \eqref{neweqna}.} at $\Xi = -3$, as depicted in Fig.~\ref{fig:deviation}. The duration at which $\Xi$ remains near this value increases rapidly near $\alpha_C$. 

As one might expect, this behavior actually occurs for \textit{any point on the boundary of the basin of attraction}.

From \eqref{pertsa}, we see that the linear perturbations in this regime obey,
$$u_{k}^{\prime\prime} + \left(1-\frac{1}{2}\phi^{\prime 2}\right)u_k^{\prime} + \left[\left(\frac{k}{aH}\right)^2 + (-2)\left(-3 + \frac{1}{2}\phi^{\prime 2}\right)\right]u_k=0.$$
Since $\phi^{\prime}\propto\phi\propto \exp(-3N)$ in this regime, this rapidly approaches
\begin{equation}\label{criticalpertsa}u_{k}^{\prime\prime} + u_k^{\prime} + \left[\left(\frac{k}{aH}\right)^2 + 6\right]u_k=0.\end{equation}

For perturbations well outside the horizon --- small $k$ --- the six dominates the last term. This represents an underdamped harmonic oscillator, with frequency $\sqrt{6}$, nothing close to scale invariance. All this unstable behavior near $\alpha_C$, leads us to conjecture that the semi-classical (mean field) theory is ill-defined near this special value of the coupling. We avoid such critical points in our present work, {and shall leave its study for future investigation.}

\section{$\Xi$ and Power Spectrum}\label{sec:three}
\subsection{Derivation}
\subsubsection*{Mode Equations}
The Mukhanov-Sasaki variable, $u$ corresponds to fluctuations of the gravitational field. Upon Fourier transforming, the modes $u_k$ obey \cite{mukhanov} equations of the form,
\begin{equation}\label{ques}\frac{\partial^2u_k}{\partial \eta^2} + (k^2 - f_u)u_k = 0.\end{equation}
Here the functions $f_u$ are given by
\begin{equation}\label{effyu} f_u = -(aH)^2\left[(1+\Xi)(\Xi+\frac{1}{2}\phi^{\prime 2})-\Xi^{\prime}\right]\end{equation}
A solution to \eqref{ques} becomes simple upon making a simple ansatz,
\begin{equation}u_k = \sqrt{\eta}Z_{M}.\end{equation}
This reduces to an equation for $Z_{M}$,
\begin{equation}\label{modeqn}\frac{\partial^2Z_M}{\partial\eta^2} + \frac{1}{\eta}\frac{\partial Z_M}{\partial\eta} + \left[k^2-\left(f_Q + \frac{1}{4\eta^2}\right)\right]Z_M = 0.\end{equation}
Upon making the identification,
\begin{equation}\label{emm}M^2 = \eta^2 f_u + \frac{1}{4},\end{equation}
the functions $Z_M$ solve Bessel's equation. We now apply these solutions to the study of the power spectrum of density perturbations.
\subsubsection*{Power Spectrum}
The comoving distance to the last scattering surface, $r_{\star}\sim 14$ Gpc, sets the scale of interest for the power spectrum. $\eta_k$ is the conformal time when the modes $u_k$ leave the horizon,
$$k \sim aH\big|_{\eta = \eta_k}.$$
$\eta_{\star}$ is the time when the modes associated to the scale $r_{\star}$ left the horizon. Thus, those mode which leaves before $\eta_{\star}$ have already frozen out. From that fact we can write the most general solution to \eqref{modeqn},
\begin{equation}\label{eqn:mode}u_k = A(k)\sqrt{k(\eta_{\star}-\eta_k)}H^{(1)}_M(k(\eta_{\star}-\eta_k))+B(k)\sqrt{k(\eta_{\star}-\eta_k)}H^{(2)}_M(k(\eta_{\star}-\eta_k)).\end{equation}
Where the $H^{(i)}_{M}$ are Hankel functions (not to be confused with the Hubble parameter), and $M$ is defined by \eqref{emm}.

The power spectrum is proportional to the squared modulus of $u_k$, evaluated at $\eta_{\star}$,
\begin{equation}\mathcal{P}(k) = \frac{72}{25}\phi^{\prime 2}H^2|u_k|^2,\end{equation}
and is traditionally presented in terms harmonic modes observable in the sky, the angular power spectrum,
\begin{equation}\label{cs}C_{\ell} = \frac{2}{25\pi}\int_{0}^{\infty} dk k^2 \mathcal{P}(k)j_{\ell}^2(kr_{\star}).\end{equation}

\subsection{Slow Roll and Its Deviations}

Let us examine the dependence of $\mathcal{P}(k)$ on $\Xi$ more closely. From \eqref{emm} and \eqref{effyu}, one finds that the indices of the relevant Hankel functions are,
\begin{equation}\label{hanki}M = \pm i\sqrt{([\eta_{\star}-\eta_k] aH)^2\left[(1+\Xi)(\Xi+\frac{1}{2}\phi^{\prime 2})-\Xi^{\prime}\right] - \frac{1}{4}}. \end{equation}
Deep in the slow roll regime, the contribution from $f_u$ nearly vanishes, leaving $M\approx\pm1/2$. At these special values, spherical Hankel functions emerge,
$$\sqrt{\eta}H^{(i)}_{1/2}(\eta)\mapsto \sqrt{\frac{2}{\pi}}\eta \;h^{(i)}_{0}(\eta).$$
These are just plane waves, $e^{\pm i\eta}$. As such, when integrated over to find $C_{\ell}$, one finds a flat spectrum, that is, for all $\ell$,
$$\frac{\ell(\ell + 1)C_{\ell}}{2C_1}=1,$$
consistent with the Harrison-Zeldovich spectrum. 

\begin{figure}[h]
  \centering
    \includegraphics[width=0.55\textwidth]{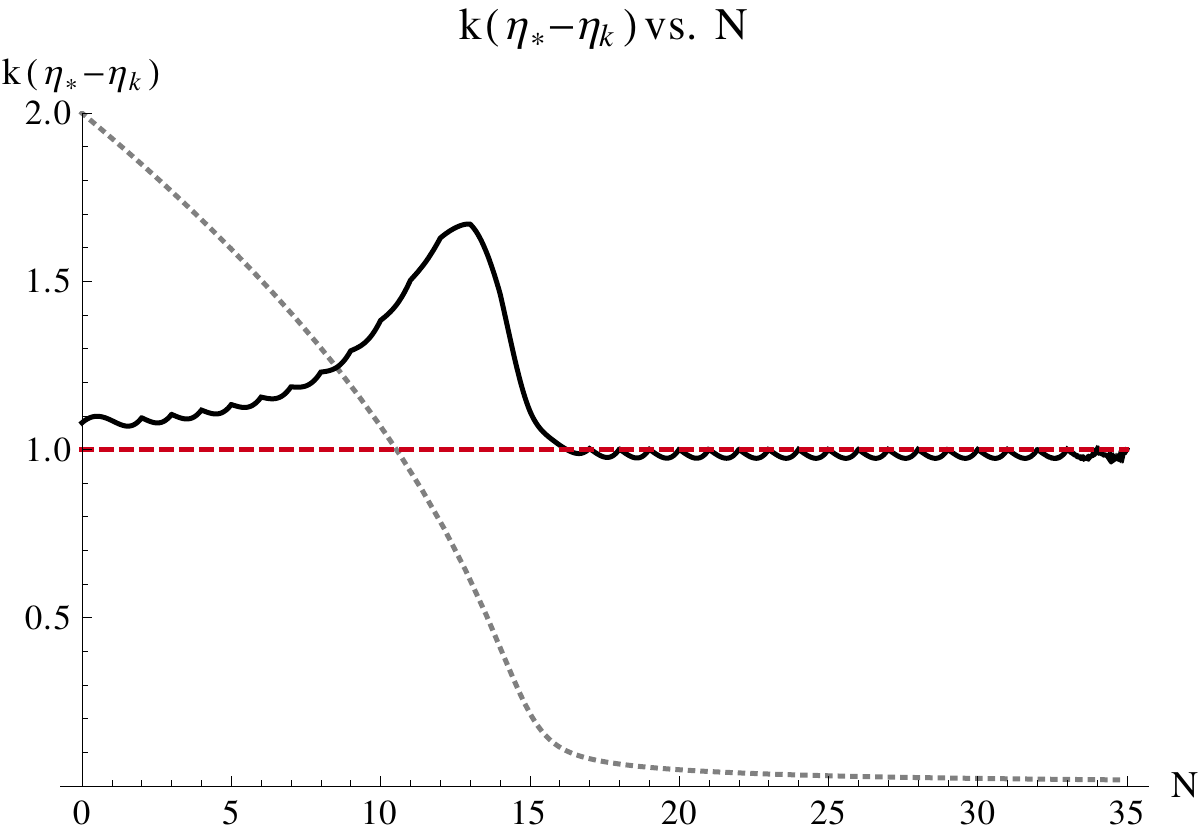}
  \caption{$x_k$ plotted against the number of e-foldings (solid black curve). During slow-roll inflation, it asymptotes to unity. Deviation from slow-roll yields a spike in this quantity. $\phi(N)$ is plotted (dashed, with arbitrary units) for reference.}
\label{fig:keta}
\end{figure}

The physical power spectrum depends on the trajectory of the inflaton. To linear order, each mode evolves independently. The power spectrum is a slice through the set of the modes at a fixed conformal time $\eta_{\star}$. Features in the trajectory of $\phi$ which occur after $\eta_{\star}$ will be imprinted on the power spectrum. To understand this, we deconstruct the quantity $x_k$, which we define as the argument of the linear perturbations \eqref{eqn:mode}, 
\begin{equation}\label{exx}x_k = \left(\frac{\partial\log(\eta_{\star}-\eta_k)}{\partial N}\right)^{-1}= k(\eta_{\star}-\eta_k).\end{equation}
That is, the modes are,
$$u_{k} = A_k \sqrt{x_k}H_{M}^{(1)}(x_k) + B_k \sqrt{x_k}H_{M}^{(2)}(x_k),$$
with 
$$ M = \pm i\sqrt{(x_k^2\left[(1+\Xi)(\Xi+\frac{1}{2}\phi^{\prime 2})-\Xi^{\prime}\right] - \frac{1}{4}}.$$
A transition from chaotic to inflection point inflation, shown in Fig.~\ref{fig:xiinflect}, can distorts $x_k$, as illustrated in Fig.~\ref{fig:keta}.

Features like that that happen at conformal time $\tilde{\eta}$ can appear in the power spectrum at $\tilde{k}\sim aH\Big|_{\eta = \tilde{\eta}}$. To be observed, such features should occur shortly before the the modes relevant for the CMB left the horizon. If the feature occurs too early, it will be pushed too far to the red part of the spectrum to be measured. This may require a fortuitous coincidence, but such signatures -- low power at large scales -- are observed in the CMB power spectrum. In any case, we now turn to  quantifying these possibilities by examining parameters and the initial conditions.

We now discuss how these features yield such a reduction of power. The important, $x_k$ dependent-quantity for this analysis is $M$, the index of the Hankel function, \eqref{hanki}. As previously mentioned, during slow-roll, $M$ asymptotes towards $1/2$. A spike in both $\Xi$ and $x_k$, as happens when $\phi$ approaches the inflection point, temporarily pushes $M$ through zero and into imaginary values. This dramatically alters the behavior of absolute value of the mode wavefunctions, $|u_k|^2$.

\begin{figure}[h]
  \centering
   \includegraphics[width=0.65\textwidth]{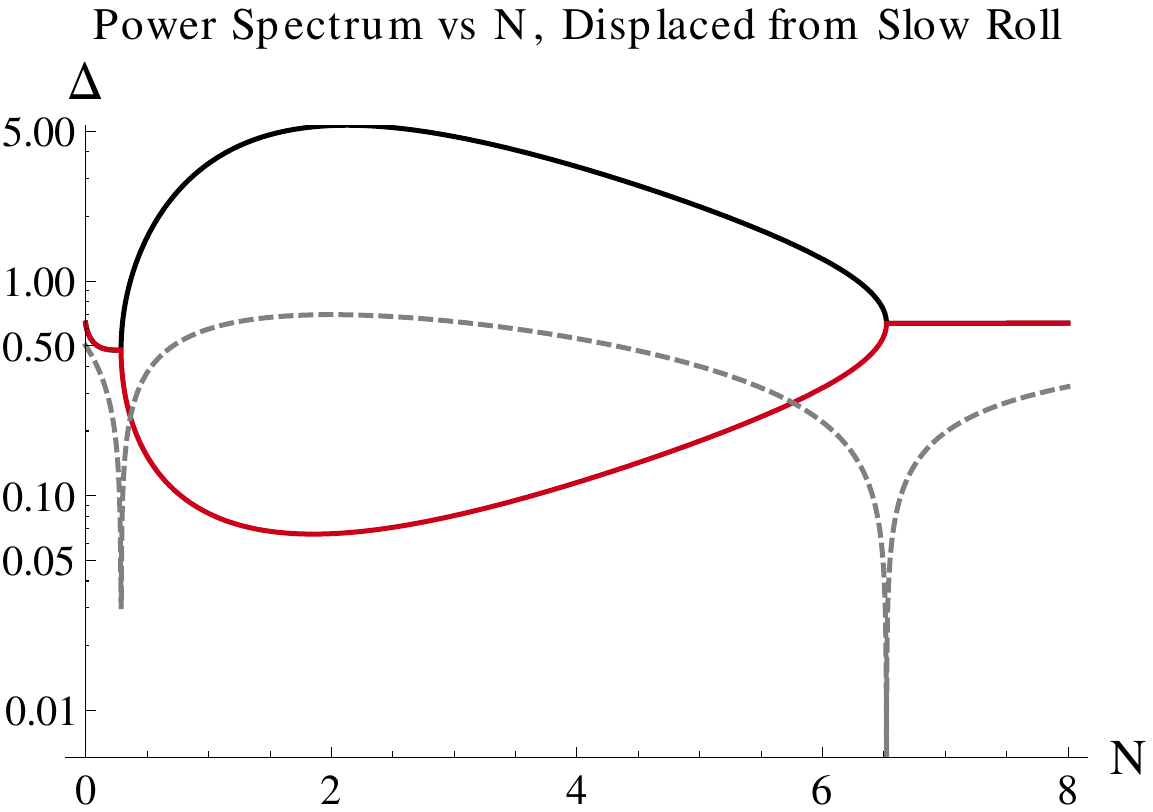}
  \caption{The power spectrum (in arbitrary units) during a spike in $\Xi$. Deviation from slow-roll yields a spike in this quantity. When separated, the red curve corresponds to the $H^{(2)}$ mode, the black is $H^{(1)}$. The magnitude of the Hankel function index, $|M|$ is plotted for reference.}
\label{fig:sobeau}
\end{figure} 

As illustrated in Fig.~\ref{fig:sobeau}, the absolute values of $\sqrt{x}H^{(1)}_M(x)$ and $\sqrt{x}H^{(2)}_M(x)$ are identical for real $M$. When $M$ passes to imaginary values, they ramify, with $\sqrt{x}H^{(1)}_{M}(x)$ exponentially growing and $\sqrt{x}H^{(2)}_{M}(x)$ decaying. This fact motivates the choice $A_{k}=0$ in \eqref{eqn:mode}, consistent with a Bunch-Davies vacuum \cite{Bunch:1978yq}.

When the transition is over, $\phi$ finds itself in slow-roll, $M$ again approaches $1/2$, and the wavefuctions resume their standard $1/k^3$ behavior. The upshot of all this is dramatic reduction of power, demonstrated by the huge ``bight'' in power spectrum during the transition. Such a feature is manifest in the red curve of Fig.~\ref{fig:sobeau}. As we shall now see, this ``bight'' has direct implications for the angular power spectrum.

\subsection{Scanning the basin of attraction}

We now quantify the effect of a transient deviation from slow-roll inflation on the power spectrum of primordial density perturbations. In particular, we focus on the effect of a transition from chaotic to inflection point inflation. We demonstrate this by examining the angular power spectrum, $C_{\ell}$, as a function of the catastrophe parameter $\alpha$ in the $A_3$ model (see Sec.~\ref{sec:basin2}).
\subsubsection*{The set up}
First we connect the theoretical ideas discussed so far to observational quantities. In particular, we need to establish the various scales in the system to carryout a numerical analysis. To that end, we begin with the normalization of the power spectrum, $\Delta^2$, which is related to the wavefunction of the linear modes \cite{mukhanov} by
$$\Delta^2 = 4k^3 H^2\phi^{\prime 2}|u_{k}|^2.$$
The $C_{\ell}$'s from Eqn.~\eqref{cs} can be reparametrized using the fact that $k=aH$, so that
\begin{equation}\label{csN}C_{\ell} = \frac{2}{25\pi}\int dN \left(1-\frac{1}{2}\phi^{\prime 2}\right) j_{\ell}^2(k(N)r_{\star}) \Delta^2(k(N)).\end{equation}
The argument of $j_{\ell}$ is dimensionless. For generic choice of units, it is given by
$$kr_{\star}\rightarrow \frac{kr_{\star}}{hc}.$$
In the vicinity of the inflection point, 
$$V \sim \frac{27}{4}\alpha^4 V_0 M_P^4,\quad \phi^{\prime}\approx 0.$$
Here $V_0$ represents the overall scaling of the potential, which though largely irrelevant for the background dynamics (c.f. Eqn.~\eqref{neweqna}), is important for the perturbations.

Simple algebra reveals that
$$kr_{\star}\rightarrow V_0^{1/2}\alpha^2 a\frac{24M_Pc^2\,\mathrm{Gpc}}{hc}\approx (V_0^{1/2}\alpha^2 a)(1.45\times 10^{60}) \approx V_0^{1/2}\alpha^2 e^{(N_k-N_0)+139},$$
where $N_0$ normalizes the scale factor to unity today.

Therefore, if the transition occurred at $N_k \sim 15$, as in Fig.~\ref{fig:xiinflect}, observability requires
\begin{equation}\label{eqn:betas} V_0^{1/2} \approx e^{N_0-124}.\end{equation}
\subsubsection*{Low power at large scales}

The primordial power spectrum appears to be very close to scale invariant. The sharp peaks and troughs observed in the CMB are well explained by baryon acoustic oscillations \cite{Eisenstein:2005su}. Power at the very largest scales, however, appears to be anomalously low. Taken together, this suggests that the universe inflated fairly regularly from $\eta_{\star}$ until the end of inflation. Any nontrivial dynamics must -- and could -- have occurred just prior to $\eta_{\star}$.

For definiteness, we focus on the transition from chaotic to inflection point inflation, although similar arguments apply for similar features. The suppression of power at low scales depends on two quantities. The time between the transition and $\eta_{\star}$ and the strength of the cubic coupling during inflation, $\alpha$. The former quantity is related to $V_0$ and the \textit{total} number of e-foldings, as can be inferred from \eqref{eqn:betas}. We define an effective parameter $\beta$ to absorb these dependencies.
$$kr_{\star}\rightarrow \beta k,$$
that is,
\begin{equation}\label{betadef}\beta = \frac{V_0^{1/2}M_Pc^2 r_{\star}}{a_0hc}.\end{equation}
Increasing $\beta$ corresponds to either raising the scale of inflation or lowering the initial total number of e-foldings, i.e. $N_0 = \log a_0$. This can be related to the initial conditions for $\phi$, and therefore the basin of attraction. 

$\alpha$ and $\beta$ have qualitatively different effects on the angular power spectrum. In particular, the effects of $\alpha$ are largely independent of $N_0$ --- the total number of e-foldings of inflation. We now study the effects of varying both $\alpha$ and $\beta$ in detail.

\subsubsection*{Changing $\alpha$ and $\beta$}
 
\begin{figure}[h]
  \centering
   \includegraphics[width=0.65\textwidth]{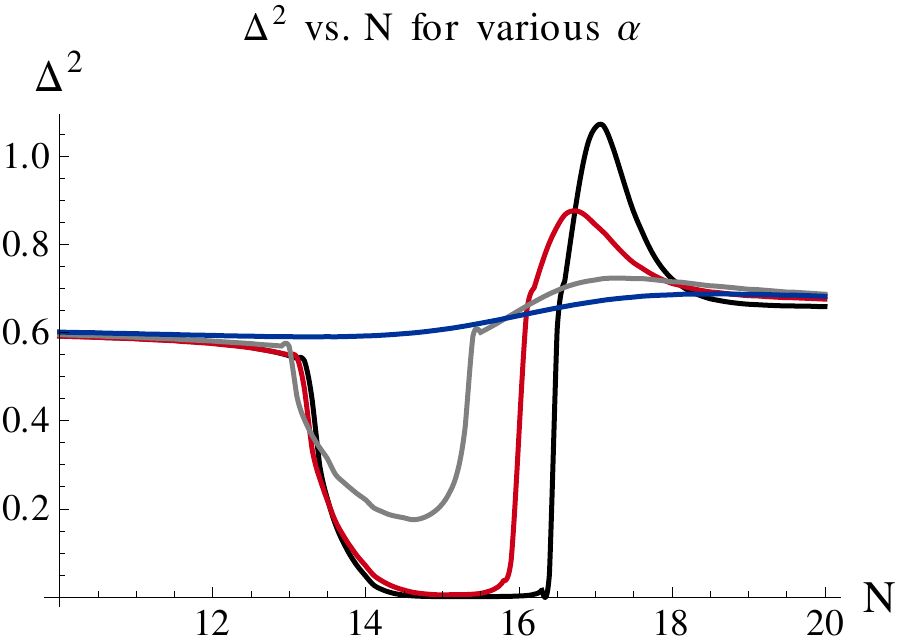}
  \caption{The power spectrum $\alpha=0.7,0.8,1.2$ and $1.6$. As $\alpha$ approaches the critical value $\alpha_c\approx0.659$, the effect on the power spectrum is more pronounced. }
\label{fig:delta}
\end{figure}

\begin{figure}[h]
  \centering
   \includegraphics[width=0.65\textwidth]{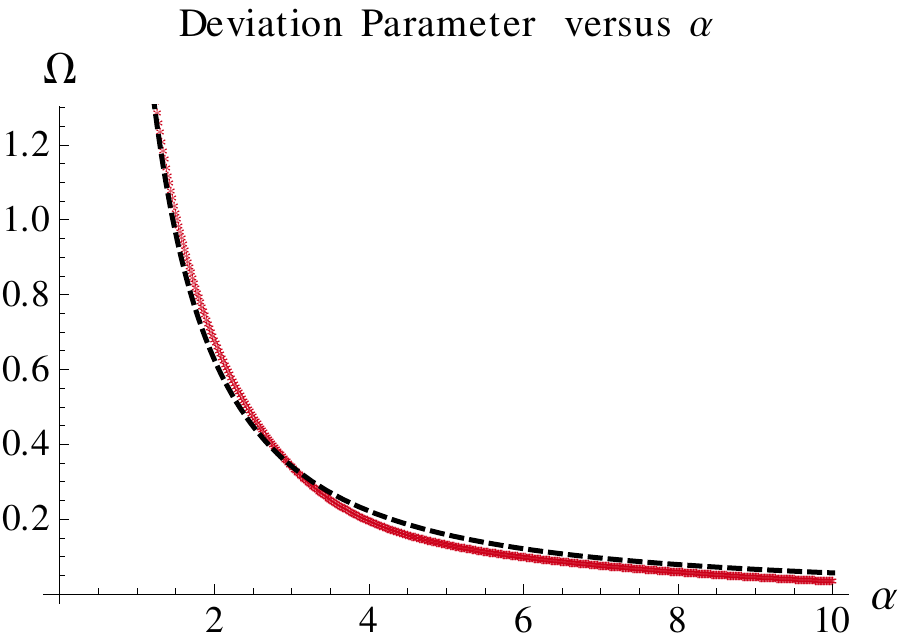}
  \caption{The slow-roll derivation parameter $\Omega$ as a function of $\alpha$ (solid red curve), the cubic coupling during inflation. $\Omega$ diverges near $\alpha_C$ and decreases as $\alpha$ grows. The dashed line represents a power law $1/N^{1.5}$. }
\label{fig:alphas}
\end{figure}

We begin by considering changes in $\alpha$. As expected from \eqref{eqn:mode}, the ``bight'' in $\Delta$ tracks the deviation of $\Xi$ from zero. Similarly, their magnitudes are correlated, as one can see by comparing Fig.~\ref{fig:inflection} and Fig.~\ref{fig:delta}. The $\alpha$ dependence of a nonzero $\Xi$ is somewhat clearer to analyze, and we do so by defining a deviation parameter $\Omega$,
\begin{equation}\label{eqn:omega}\Omega = \int \Xi^2\; dN.\end{equation}
$\Omega$ rapidly falls to zero as $\alpha$ deviates from $\alpha_C\approx 0.659$. This behavior is plotted by the solid red curve in Fig.~\ref{fig:alphas}. The integral is taken over the feature associated to the transition. For example, if the transition occurs at $N=15$, we integrate from around $N=5$ to around $N=30$, since both endpoints are well within the slow-roll regime. Of course, such a transition can occur at any value of $N$. A larger value of $\Omega$ corresponds to a more violent and sustained deviation from slow-roll and a larger ``bight'' in the power spectrum. $\Omega$ has an approximate power law dependence on $\alpha$, 
$$\Omega \propto \alpha^{-3/2},$$
which is represented by the dashed curve in Fig.~\ref{fig:alphas}. The actual scaling $\alpha$ dependence oscillates slowly between powers of $-1.8$ and $-1.3$, but the qualitative behavior is clear.

\begin{figure}[h]
  \centering
   \includegraphics[width=0.65\textwidth]{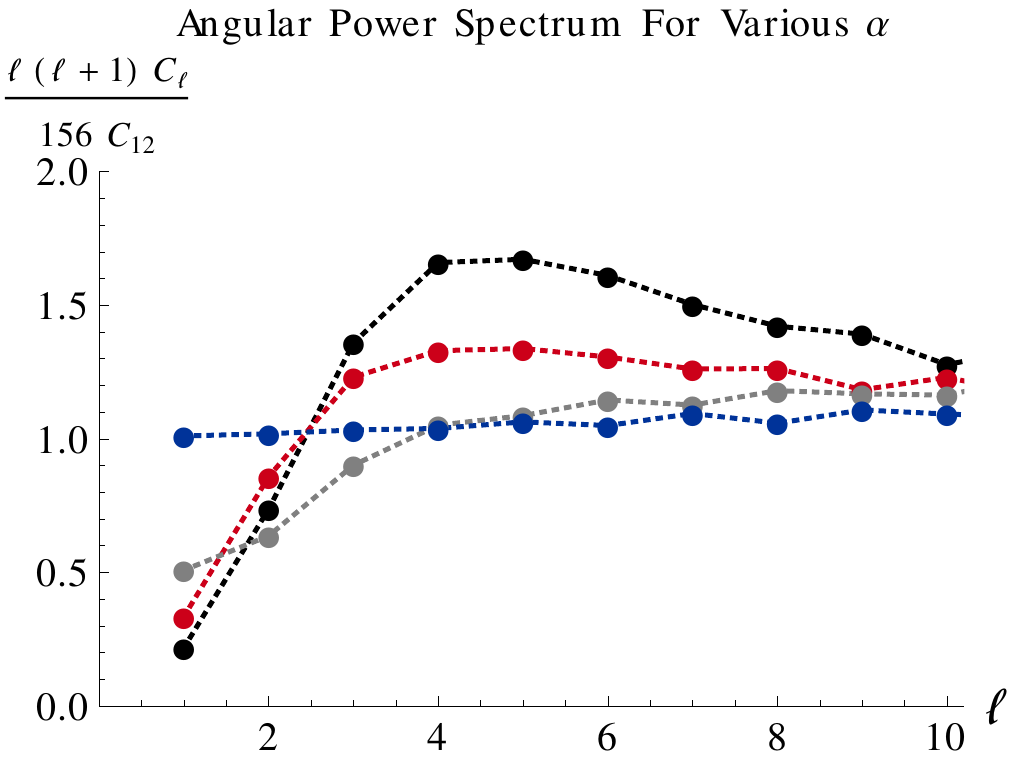}
  \caption{The angular power spectrum $\alpha=0.7,0.8,1.2$ and $1.6$. As $\alpha$ approaches the critical value $\alpha_c\approx0.659$, the effect on the power spectrum is more pronounced. }
\label{fig:reader}
\end{figure}

The temporary reduction in $\Delta$ can also be seen in the angular power spectrum. The relation between them can be seen by comparing  Fig.~\ref{fig:delta} and Fig.~\ref{fig:reader}. When these features in $\Delta$ occur on the largest observable scales, the the first few modes of the angular power spectrum are suppressed. Fig.~\ref{fig:reader} illustrates this for $\alpha = 0.7,0.8,1.2$ and $1.6$. The largest effect occurs for $\alpha=0.7$, which is close to $\alpha_C$. The values of $\alpha$ (and colors) are linked across both of these plots.

\begin{figure}[h]
        \centering
        \begin{subfigure}[b]{0.45\textwidth}
                \centering
                \includegraphics[width=\textwidth]{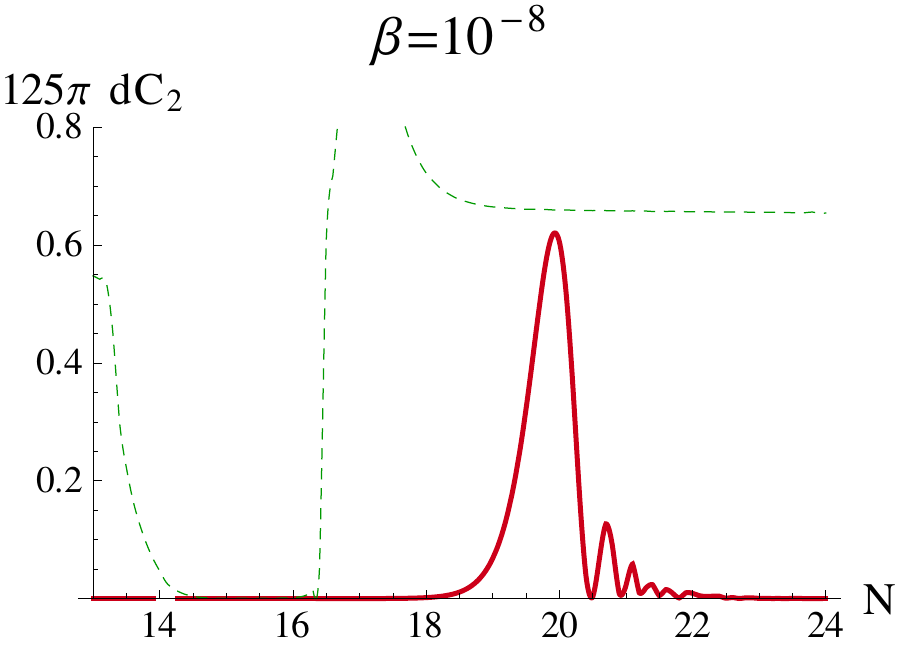}
                \caption{ }
	\label{fig:betaA}
        \end{subfigure}%
        ~ %add desired spacing between images, e. g. ~, \quad, \qquad etc. 
        \begin{subfigure}[b]{0.45\textwidth}
                \centering
                \includegraphics[width=\textwidth]{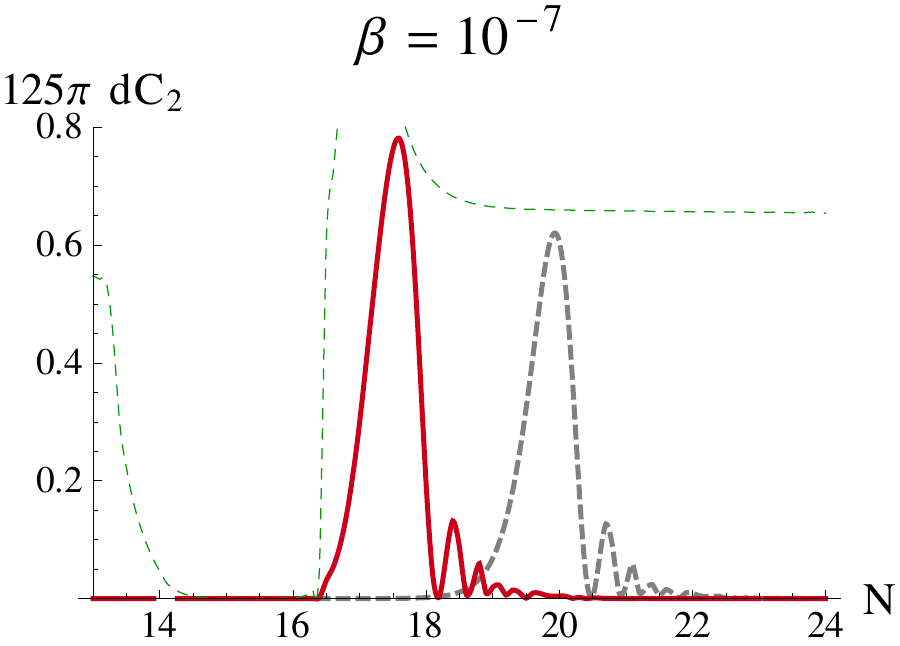}
                \caption{}
	\label{fig:betaB}
        \end{subfigure}\vskip2ex
        ~ %add desired spacing between images, e. g. ~, \quad, \qquad etc. 
        \begin{subfigure}[b]{0.45\textwidth}
                \centering
                \includegraphics[width=\textwidth]{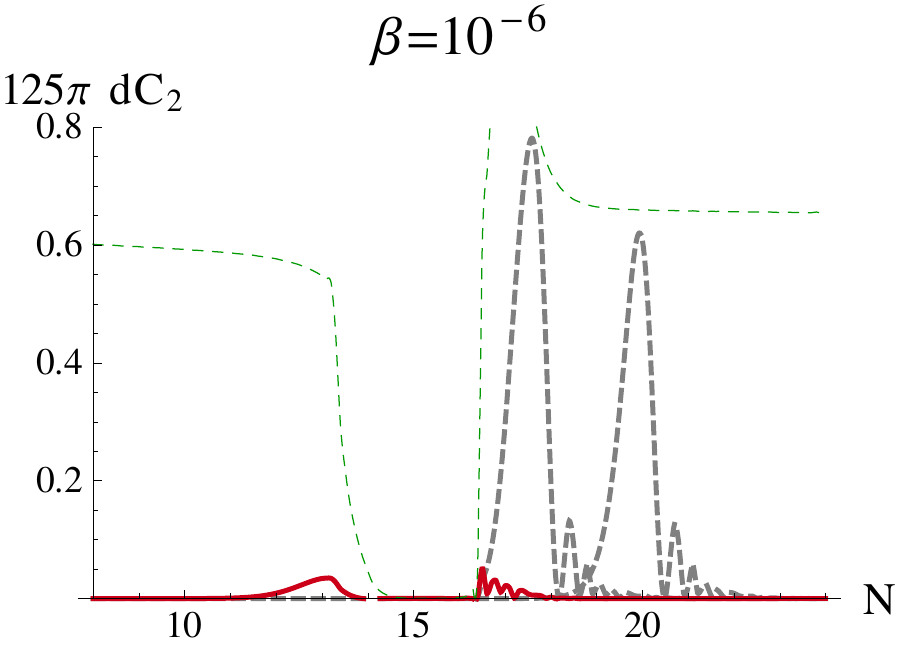}
                \caption{ }
		\label{fig:betaC}
        \end{subfigure}
        \begin{subfigure}[b]{0.45\textwidth}
                \centering
                \includegraphics[width=\textwidth]{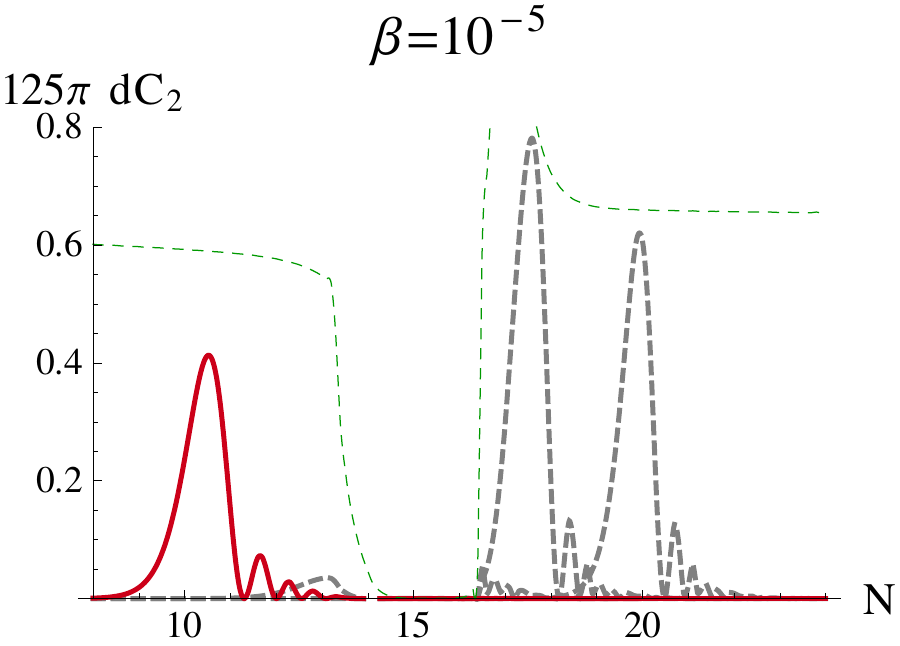}
              \caption{ }
	\label{fig:betaD}
        \end{subfigure}

        \caption{Integrand of the Quadrupole moment, $dC_{2}$ versus $N$, for various values of $\beta$, defined in Eqn.~\ref{betadef}. The solid red curve corresponds to the $\beta$ indicated in each panel. The grey dashed curves are the $dC_{2}$ from previous panels, replotted for reference. As $\beta$ increases, $dC_2$ shifts left. $\Delta^2$, which acts as an envelope, is also plotted for reference (thin, dashed curve). We chose $\alpha=0.7$ for this figure.}\label{fig:betas}
\end{figure}

\vskip2ex
We now turn to $\beta$, which determines which range of $N$ dominate the angular power spectrum --- the $C_{\ell}$ integrals. The integrands, $dC_{\ell}$, are the oscillating curves in Fig.~\ref{fig:betas}. Large values of $\beta$ shifts the curves to the left --- to larger scales. These integrals are enveloped by $\Delta^2$, which \textit{does not} change with $\beta$. These are the light, dashed curves seen in Fig.~\ref{fig:betas}. As we just discussed, $\alpha$ changes the shape of the envelope. In particular, close to the critical value, $\alpha_C$, a larger ``bight'' is removed from the envelope.

%%%%%%%%%%%%%%%%%%%%%%%%%%%%%%%%%%%%%%%%%%%%%%%%
%%%%%%%%%%%%%%%%%%%%%%%%%%%%%%%%%%%%%%

To explain the anomaly in the CMB angular power spectrum, $\beta$ must take a value close to that in Fig.~\ref{fig:betaC}. That is, the associated power reduction must be focused on the low multipole moments. Since no such power reduction appears seems to exist at smaller scales, to be physical it may also be smaller. While it is fixed by these considerations, from \eqref{betadef} we see that there is still a degenerancy between $V_0$ and $a_0$. Since $a_0$ depends on the total expansion of the universe before today, it can be related to the initial conditions. In particular, for fixed $V_0$ we can see how $\beta$ varies over the entire basin of attraction.

\begin{figure}[h]
        \centering
        \begin{subfigure}[b]{0.48\textwidth}
                \centering
                \includegraphics[width=\textwidth]{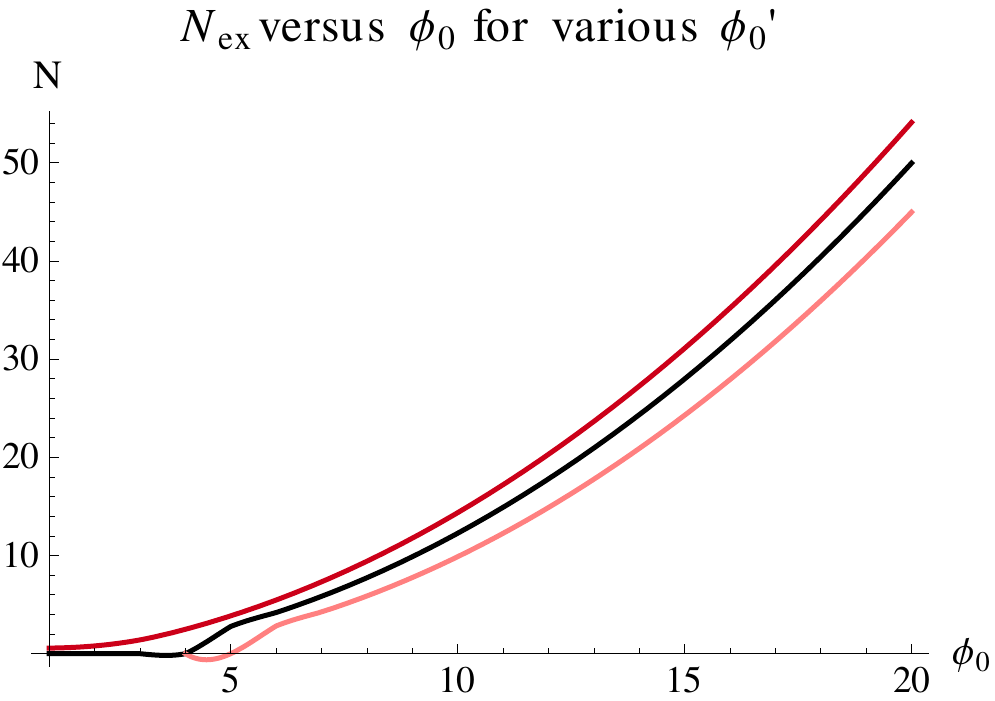}
                \caption{Curves of constant initial velocity}
	\label{fig:xxA}
        \end{subfigure}%
        ~ %add desired spacing between images, e. g. ~, \quad, \qquad etc. 
        \begin{subfigure}[b]{0.5\textwidth}
                \centering
                \includegraphics[width=\textwidth]{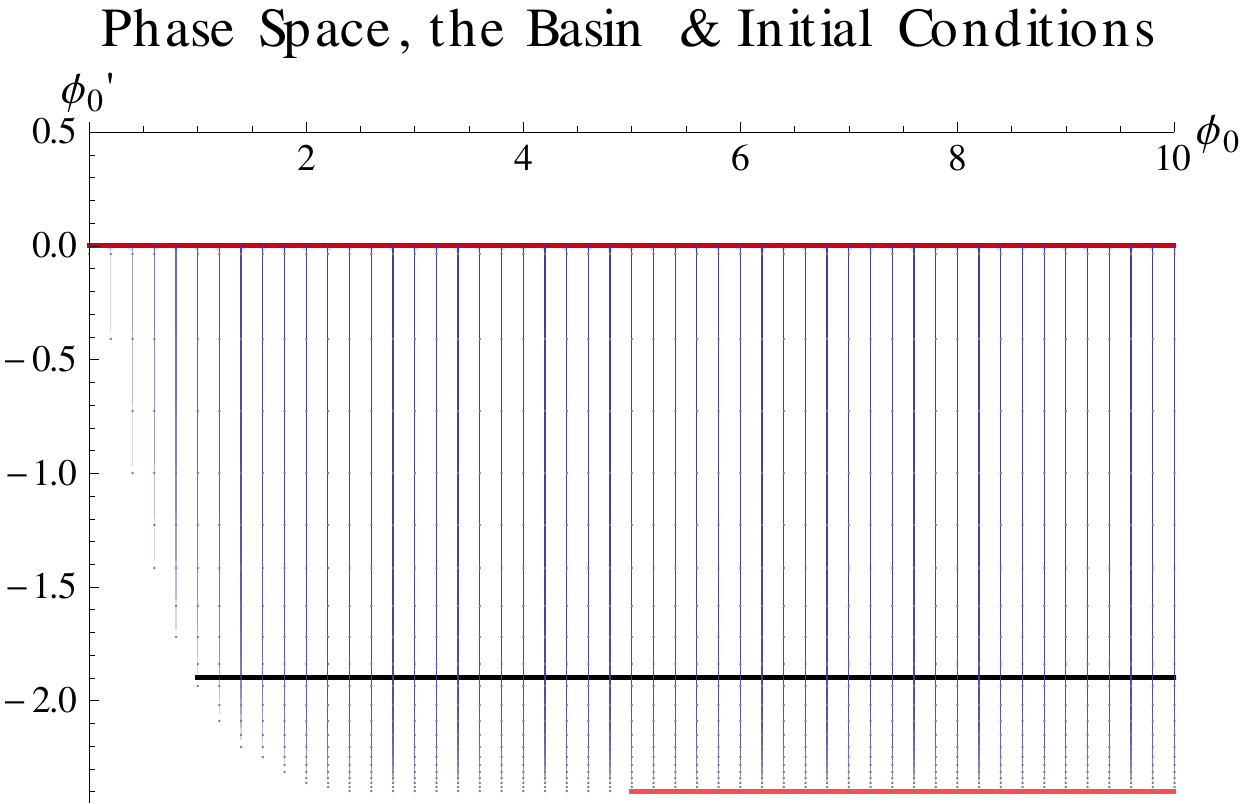}
              \caption{Curves of constant initial velocity within the basin of attraction.}
	\label{fig:xxB}
        \end{subfigure}

        \caption{The basin of attraction is plotted with curves of constant initial $\phi^{\prime}$. The red, black and orange curves correspond to $\phi^{\prime} = 0,-1.9,-2.4$, respectively. Deeper into the basin, the more e-foldings of inflation occur above the inflection point. Note how little initial $\phi^{\prime}$ suppresses $N_{ex}$. We chose $\alpha=1$ in this figure.}\label{fig:xx}
\end{figure}

The amount of inflection point inflation depends on $\lambda$, and therefore independent of the initial conditions. However, the length and duration of the approach to the inflection point gives rise to a prior history of chaotic inflation. Call this ``extra'' inflation $N_{ex}$. It scales by $\beta$ by 
$$\beta\rightarrow \exp(-N_{ex})\beta.$$
This is the effect, illustrated in Fig.~\ref{fig:xx}, which we now quantify. Any initial field velocity will drastically reduce $N_{ex}$, but starting from a larger value of $\phi$ will increase it. Each curve in both Fig.~\ref{fig:xxA} and Fig.~\ref{fig:xxB} corresponds to a different initial velocity: $\phi^{\prime}=0,-1.9,-2.4$. These are the red, black and orange curves, respectively. Note that the maximum possible $\phi^{\prime}$ is $-\sqrt{6}\sim -2.449$, as this corresponds to kinetic energy domination. In Fig.~\ref{fig:xxB}, the basin of attraction is sketched for reference. The important effect to observe is that $N_{ex}$ grows rapidly as the initial conditions are chosen deeper inside the basin of attraction. The larger velocities are have little effect on $N_{ex}$, as can be seen in Fig.~\ref{fig:xxA}. Indeed, $N_{ex}$ is only strongly suppressed exponentially close to the boundary of the basin of attraction, which is shown in Fig.~\ref{fig:basin}.

\subsubsection*{Implications for the measure problem}
A number of studies have attempted to put a measure on phase space. The standard Liouville measure suggests that early-time kinetic domination is by far the most likely scenario. While including the couplings amongst the random variables enhances the likelihood of inflection point inflation to $1/N^3$, this measure still favors a substantial initial velocity. This suggests that a larger $\beta$ is far more likely. In short, $N_{ex}\sim 0$. In this case \eqref{betadef} together with \eqref{efolds} gives,
$$\beta \approx \sqrt{V_0}\exp(-\pi/2\sqrt{3\alpha \lambda})\frac{M_Pc^2r_{\star}}{hc}.$$
For $\beta$ fixed by the CMB, and fixed $\alpha$, the scale of inflection point inflation is fixed by its duration,
$$V_0\propto e^{2/\sqrt{\lambda}}.$$
Note that $\lambda$ must be sufficiently small to generate sufficient expansion for the observable universe. We also stress here that this analysis is extremely sensitive to the details of the chosen measure on phase space, although it is interesting to see further implications of the measure used in \cite{Gibbons:2006pa}. More generally, one must appeal to Fig.~\ref{fig:xx}.

\section{Conclusion}\label{sec:four}

The dynamics of the inflaton have a rich structure despite the generic predictions for the cosmic microwave background. Advances in observational and theoretical technologies have increased our sensitivity to the effects of nontrivial dynamics on the cosmological perturbations. Small-field models of inflation generically involve temporary deviations from the slow-roll, attractor trajectory. In this work we have quantified these deviations and shown how they may affect the primordial density perturbations. Crucially, these effects arise in a model independent fashion. Since these effects only depend on the local structure of the potential, they scale with the same universality parameters discussed in early work \cite{Downes:2011gi}.

We demonstrated analytically how sufficiently large deviations from slow-roll change the structure of the wavefunction for the perturbations, and explicitly how this reduced the power in their spectrum. We closed by relating both the couplings and the initial conditions to the strength and timing of this power suppression. While this completes the systematics for how low power at large scales may have arisen with a chosen universality class, we also discussed how it informs the likelihood of inflation.

This sudden change in the power spectrum may have important implications for nongaussianities, particularly in the context of multifield models where the dynamics have a richer structure. It would be interesting understand the relation, if any, between the strength and shape of nongaussianities and the universality parameters.

Finally, the deviation from slow-roll is most dramatic near critical values of the couplings. This fact leads to a curious saturation of the $\Xi$ parameter, and has an extreme effect on the perturbations. Itzhaki and Kovetz \cite{Itzhaki:2008hs} showed that the background has the properties of a second order phase transition. More generally, this feature occurs along the entire boundary of the basin of attraction. As $\Xi$ is maximized for an extended period, it stands to give the strongest effect on primordial nongaussianities. From a field theoretic perspective, such a dramatic effect near the vicinity of a nontrivial fixed point alone warrants further investigation.

\section*{Acknowledgements}
The authors wish to thank Eiichiro Komatsu, Ely Kovetz and Todd Zapata for helpful discussions. S.D. would also like to thank the Physics department at Northeastern University for their hospitality where parts of this work was completed. This work is supported in part by the DOE grant DE-FG02-95ER40917 and the Mitchell Institute for Fundamental Physics and Astronomy.
\bibliography{bibPerts}{}

\end{document}